\documentclass[aps,prb,10pt, twocolumn, superscriptaddress, longbibliography, nobalancelastpage]{revtex4-1}

\usepackage{amsmath}
\usepackage{amssymb}
\usepackage{graphicx}
\usepackage{dsfont}
\usepackage{braket}
\usepackage{bm}
\usepackage{wrapfig}
\usepackage{kantlipsum}
\usepackage{mathtools}
\usepackage[colorlinks=true,linkcolor=red,urlcolor=magenta,citecolor=blue]{hyperref}
\usepackage{tikz}
\usepackage[english]{babel}
\usepackage[normalem]{ulem}
\usepackage{bibentry}

\definecolor{g}{rgb}{.1,0.4,.1} 
\definecolor{b}{rgb}{0,0.2,1}
\definecolor{rouge}{rgb}{0.82,0.,0.}
\definecolor{vert}{rgb}{0.,0.82,0.}
\definecolor{orange}{rgb}{1,0.5,0.}
\definecolor{bleu}{rgb}{0.,0.,0.82}
\definecolor{m}{rgb}{0.82,0.,0.82}
\definecolor{vert2}{rgb}{0.,0.5,0.}
\definecolor{rougeclair}{rgb}{1.0,0.7,0.7}
\definecolor{gris}{rgb}{.8,.8,.8} 

\newcommand{\e}{\mathrm{e}}
\renewcommand{\d}{\mathrm{d}}

\newenvironment{diagram}
{
	\begin{tikzpicture}[baseline = (X.base),every node/.style={scale=0.8},scale=.45]
}
{
	\end{tikzpicture}
}

\begin{document}

\title{Real-time scattering of interacting quasiparticles in quantum spin chains}

\author{Maarten Van Damme}
\email{maarten.vandamme@ugent.be}
\affiliation{Department of Physics and Astronomy, University of Ghent, Krijgslaan 281, 9000 Gent, Belgium}

\author{Laurens Vanderstraeten}
\affiliation{Department of Physics and Astronomy, University of Ghent, Krijgslaan 281, 9000 Gent, Belgium}

\author{Jacopo De Nardis}
\affiliation{Department of Physics and Astronomy, University of Ghent, Krijgslaan 281, 9000 Gent, Belgium}

\author{Jutho Haegeman}
\affiliation{Department of Physics and Astronomy, University of Ghent, Krijgslaan 281, 9000 Gent, Belgium}

\author{Frank Verstraete}
\affiliation{Department of Physics and Astronomy, University of Ghent, Krijgslaan 281, 9000 Gent, Belgium}

\begin{abstract}
We develop a method based on tensor networks to create localized single particle excitations on top of strongly-correlated quantum spin chains. In analogy to the problem of creating localized Wannier modes, this is achieved by optimizing the gauge freedom of momentum excitations on top of matrix product states. The corresponding wave packets propagate almost dispersionless. The time-dependent variational principle is used to scatter two such wave packets, and we extract the phase shift from the collision data. We also study reflection and transmission coefficients of a wave packet impinging on an impurity.
\end{abstract}

\maketitle

The field of strongly-correlated quantum many-body physics is one of the most interesting and diverse fields in contemporary physics, both theoretically and experimentally. Here the notion of a quasiparticle is commonly understood as carrying the low-energy degrees of freedom on top of a strongly-correlated ground state. With traditional experimental probes, these quasiparticles are targeted in the momentum-energy plane---think of inelastic neutron-scattering experiments on low-dimensional quantum magnets as an outstanding example \cite{Stone2006, Coldea2010, Mourigal2013}. Recently, however, progress in the design and probing of strongly-correlated quantum phases has made it possible to directly observe the dynamics of quasiparticles in real-space and real-time \cite{Jurcevic2014, Chiu2019, Koepsell2019, Vijayan2020}. 

\par On the theoretical side, understanding low-energy dynamics in terms of effective quasiparticles originates from Fermi liquid theory, where quasiparticles are close to their free counterparts but the interaction, treated perturbatively, leads to dressing and finite lifetimes. The notion of quasiparticles has also been made precise for integrable spin chains by the recent development of generalized hydrodynamics \cite{Bertini2016, Castro-Alvaredo2016, Bulchandani2017}, which fully describes the large-scale dynamics of a system only in terms of the semi-classical motion of stable quasiparticles. This idea can also in principle work for non-integrable systems, where local quasiparticles are usually thought of as almost stable excitations with long lifetimes that can be treated semi-classically \cite{Damle1998, Moca2017}. However, there is much less information available about the dynamics of local excitations in generic interacting systems and there is no generic procedure to extract them from microscopics.

\par Recently, a variational framework was developed for representing quasiparticles in gapped one-dimensional quantum spin systems in the thermodynamic limit, which does not rely on integrability. This framework starts from a variational ansatz \cite{jutho2012,haegeman2013post} for describing elementary excitations on top of strongly-correlated ground states parametrized by the class of matrix product states (MPS) \cite{verstraete2009, Schollwoeck2011, orus2013}.  Inspired by the single-mode approximation\cite{feynman1954, haldane1988}, the ansatz effectively boils down to a plane wave of a local perturbation on top of a strongly-correlated ground state. This approach is very reminiscent of a quasiparticle running through the system, but differs from the traditional quasiparticle concept since these states are exact eigenstates (up to variational errors) with infinite lifetimes with respect to the fully-interacting hamiltonian \cite{jutho2013}, much like electrons and protons in the standard model. In a natural next step, two-particle wave functions were constructed and a generic definition of the two-particle $S$-matrix was proposed \cite{laurens2014, laurens2015}.

In this Letter, a method is presented that yields insight in the propagation and mutual interaction of these quasiparticle excitations in real space and time. Whereas in some past works this was done for integrable systems on top of product states \cite{Vlijm2015, Ganahl2012}, our method is used to study two-particle interactions in both integrable and non-integrable systems on the correlated ground state, and is therefore truly capturing the low-energy degrees of freedom for strongly-interacting systems. We also study the scattering of a particle off an impurity and how this impedes the transport of energy.

\noindent\emph{\textbf{Methodology}}. %
As a first step, we approximate the ground state in the thermodynamic limit as a uniform matrix product state \cite{jutho2011, laurens2019} (MPS), which we can represent as 
\begin{equation} 
\ket{\Psi(A)} = 
\begin{diagram}
\draw[dotted] (0,1.5) -- (0.5,1.5); 
\draw (0.5,1.5) -- (1,1.5); 
\draw[rounded corners] (1,2) rectangle (2,1);
\draw (1.5,1.5) node (X) {$A$};
\draw (2,1.5) -- (3,1.5); 
\draw[rounded corners] (3,2) rectangle (4,1);
\draw (3.5,1.5) node {$A$};
\draw (4,1.5) -- (5,1.5);
\draw[rounded corners] (5,2) rectangle (6,1);
\draw (5.5,1.5) node {$A$};
\draw (6,1.5) -- (7,1.5); 
\draw[rounded corners] (7,2) rectangle (8,1);
\draw (7.5,1.5) node {$A$};
\draw (8,1.5) -- (9,1.5); 
\draw[rounded corners] (9,2) rectangle (10,1);
\draw (9.5,1.5) node {$A$};
\draw (10,1.5) -- (10.5,1.5);
\draw (1.5,1) -- (1.5,.5); \draw (3.5,1) -- (3.5,.5); \draw (5.5,1) -- (5.5,.5);
\draw (7.5,1) -- (7.5,.5); \draw (9.5,1) -- (9.5,.5); 
\draw[dotted] (10.5,1.5) -- (11,1.5); 
\end{diagram}\,,
\end{equation}
where every node corresponds to a tensor and every connected node is a contraction over the corresponding index. This class of states has a number of desirable properties: the ground state of generic gapped quantum spin chains can be faithfully represented as an MPS \cite{hastings2006, verstraete2006}, which allows for efficient calculation of all observables \cite{verstraete2009, Schollwoeck2011, orus2013}. Approximating the ground state of a given model hamiltonian within this manifold can be done using variational optimization methods \cite{zauner2018,laurens2019}.

\par Next, the quasiparticle excitations on this ground state can be described using the MPS quasiparticle ansatz \cite{jutho2012},
\begin{equation} 
\ket{\Phi_k(B)} = \sum_{n} \e^{ikn}\;
\begin{diagram}
\draw (0.5,1.5) -- (1,1.5); 
\draw[rounded corners] (1,2) rectangle (2,1);
\draw (1.5,1.5) node (X) {$A$};
\draw (2,1.5) -- (3,1.5); 
\draw[rounded corners] (3,2) rectangle (4,1);
\draw (3.5,1.5) node {$A$};
\draw (4,1.5) -- (5,1.5);
\draw[rounded corners] (5,2) rectangle (6,1);
\draw (5.5,1.5) node {$B$};
\draw (6,1.5) -- (7,1.5); 
\draw[rounded corners] (7,2) rectangle (8,1);
\draw (7.5,1.5) node {$A$};
\draw (8,1.5) -- (9,1.5); 
\draw[rounded corners] (9,2) rectangle (10,1);
\draw (9.5,1.5) node {$A$};
\draw (10,1.5) -- (10.5,1.5);
\draw (1.5,1) -- (1.5,.5); \draw (3.5,1) -- (3.5,.5); \draw (5.5,1) -- (5.5,.5);
\draw (7.5,1) -- (7.5,.5); \draw (9.5,1) -- (9.5,.5);
\draw (1.5,0) node {$\dots$}; \draw (3.5,0) node {$s_{n-1}$}; \draw (5.5,0) node {$s_{n}$}; \draw (7.5,0) node {$s_{n+1}$}; \draw (9.5,0) node {$\dots$};
\end{diagram} .
\end{equation}
Because the tensor $B$ acts on the virtual level of the MPS, it can capture the deformation of the ground state over an extended region, and therefore represents a dressed object on top of a correlated ground state. For a given momentum $k$, the problem of minimizing the energy can be reduced to an eigenvalue problem for the tensor $B$, which can be efficiently solved using iterative eigensolvers. The resulting solution depends on $k$ and yields the dispersion relation $\epsilon_k$ as eigenvalue, while the corresponding eigenvector is henceforth denoted as $B_k$. It was shown \cite{jutho2012, ZaunerStauber2018} that this approach yields quasi-exact results for the dispersion relation of generic spin chains. Additionally, these variational quasiparticles are experimentally relevant, as in gapped systems they carry the major fraction of the spectral weight of local operators, and they can be used to compute spectral functions to high precision \cite{Bera2017}.

\par Using these momentum-resolved quasiparticle states, we can now localize a wave packet in real space  via a procedure reminiscent to the construction of Wannier orbitals \cite{Marzari2012}. A real-space localized wave packet, centered around site $m$, is given by
\begin{align}
\ket{\Phi_\mathrm{rs}(m)} &= \int_0^{2\pi} \e^{-ikm} f(k) \ket{\Phi_k(B_k)}  \d k \\
&= \sum_n 
\begin{diagram}
\draw (0.5,1.5) -- (1,1.5); 
\draw[rounded corners] (1,2) rectangle (2,1);
\draw (1.5,1.5) node (X) {$A$};
\draw (2,1.5) -- (3,1.5); 
\draw[rounded corners] (3,2) rectangle (4,1);
\draw (3.5,1.5) node {$A$};
\draw (4,1.5) -- (5,1.5);
\draw[rounded corners] (5,2) rectangle (6,1);
\draw (5.5,1.5) node {$B_n$};
\draw (6,1.5) -- (7,1.5); 
\draw[rounded corners] (7,2) rectangle (8,1);
\draw (7.5,1.5) node {$A$};
\draw (8,1.5) -- (9,1.5); 
\draw[rounded corners] (9,2) rectangle (10,1);
\draw (9.5,1.5) node {$A$};
\draw (10,1.5) -- (10.5,1.5);
\draw (1.5,1) -- (1.5,.5); \draw (3.5,1) -- (3.5,.5); \draw (5.5,1) -- (5.5,.5);
\draw (7.5,1) -- (7.5,.5); \draw (9.5,1) -- (9.5,.5);
\draw (1.5,0) node {$\dots$}; \draw (3.5,0) node {$s_{n-1}$}; \draw (5.5,0) node {$s_{n}$}; \draw (7.5,0) node {$s_{n+1}$}; \draw (9.5,0) node {$\dots$};
\end{diagram} \;,
\end{align}
where the real-space tensor $B_n$ is
\begin{equation}
B_n = \int_0^{2\pi} f(k) B_k e^{ik(n-m)} dk.
\end{equation}

The determination of $B_n$ is ambiguous in three ways. First, since the tensors $B_k$ are found as a solution of an eigenvalue problem, they are only defined up to a phase, just as in the original Wannier orbital problem, and the sampling function $f(k)$ should compensate for this. But there are additional $k$-dependent gauge redundancies in the tensor $B_k$, since the transformation
\begin{equation}
B_k \to B_k + e^{ik} A X_k - X_k A
\end{equation}
with $X$ a matrix that multiplies the MPS tensor $A$ on the virtual level, gives rise to the same wave function $\ket{\Phi_k(B_k)}$. A third difficulty concerns the fact that, in a practical computer simulation, we only have access to a finite set of momenta, such that the real-space tensor
\begin{equation}
B_n = \sum_{j=1}^N f(k_j) B_{k_j} e^{ik_jn}, \qquad k_j = j \frac{2\pi}{N}
\end{equation}
is periodic with $n\to n+N$ and thus contains an infinite number of translated copies of the wave packet.
\par We first resolve the gauge freedom by tuning $X_k$ for the tensor $B_k$, so as to make its effect well localized and symmetric. A natural choice is to minimize the matrix norm of 
\begin{equation}
\begin{diagram}
\draw (0,.5) node (X) {$\phantom{X}$};
\draw[rounded corners] (1,2) rectangle (2,1);
\draw (1.5,1.5) node {$B_k$};
\draw[rounded corners] (1,-1) rectangle (2,0);
\draw (1.5,-.5) node {$A_C$};
\draw[rounded corners] (-0.5,2) rectangle (0.5,1);
\draw (0,1.5) node {$A_L$};
\draw[rounded corners] (-0.5,-1) rectangle (0.5,0);
\draw (0,-.5) node {$A_L$};
\draw[rounded corners] (2.5,2) rectangle (3.5,1);
\draw (3,1.5) node {$A_R$};
\draw[rounded corners] (2.5,-1) rectangle (3.5,0);
\draw (3,-.5) node {$A_R$};
\draw (2,1.5) -- (2.5,1.5); \draw (0.5,1.5) -- (1,1.5); \draw (-1,1.5) -- (-0.5,1.5);
\draw (-1,-.5) -- (-0.5,-.5); \draw (3.5,-.5) -- (4.0,-.5);
\draw (2,-.5) -- (2.5,-.5); \draw (0.5,-.5) -- (1,-.5);
\draw (1.5,1) -- (1.5,0); \draw (-0,1) -- (-0,0); 
\draw (3,1) -- (3,0); \draw (3.5,1.5) --(4.0,1.5);
\end{diagram}\; ,
\end{equation}
which ensures that the perturbation of the ground state by the tensor is as local as possible\cite{suppl}; here, we have used the `center-gauged' MPS tensors \cite{laurens2019} $\{A_L,A_R,A_C\}$ in order to orthogonalize the state locally (tracing on the left and right in the above tensor yields zero). Next, we find the sampling function $f(k)$ such that $B_n$ is confined to the smallest possible region in space. For an odd number of sampling points $N = 2L + 1$, we find the minimal length $\Lambda < L $ for which there still exists an $f(k)$ such that
\begin{equation}
\left| B_n \right| < \epsilon, \quad \forall n \in [-L,-\Lambda] \cup [+\Lambda,+L]
\end{equation}
for some chosen small parameter $\epsilon$ \cite{suppl}. 
Because the corresponding $B_n$ is then localised within the region $[-\Lambda,\Lambda]$, we can safely truncate the shifted copies by convoluting with a filter function, and as such obtain the definition of our wave packet. Finally, we then prepare wave packets centered around a given momentum $k_0$ by multiplying the sampling function $f$ from the previous construction with an additional gaussian distribution, i.e.\
\begin{equation} \label{eq:gausint}
f(k_j) \rightarrow f(k_j) e^{\frac{(k_0-k_j)^2}{2v^2}}  .
\end{equation}
This completely solves the problem of constructing a narrow wave packet with well defined momentum.

We can simulate the collision of two such wave packets centered around different momenta at different locations in real space by representing this state as a finite string of tensors $V_i$ surrounded by the uniform MPS tensors, 
\begin{multline} 
\ket{\Psi(A;V_1,...,V_n)} \\ =  
\begin{diagram}
\draw[dotted] (0,1.5) -- (0.5,1.5); 
\draw (0.5,1.5) -- (1,1.5); 
\draw[rounded corners] (1,2) rectangle (2,1);
\draw (1.5,1.5) node (X) {$A$};
\draw (2,1.5) -- (3,1.5); 
\draw[rounded corners] (3,2) rectangle (4,1);
\draw (3.5,1.5) node {$V_1$};
\draw (4,1.5) -- (4.5,1.5);
\draw[dotted] (4.5,1.5) -- (5.5,1.5);
\draw (5.5,1.5) -- (6,1.5); 
\draw[rounded corners] (6,2) rectangle (7,1);
\draw (6.5,1.5) node {$V_n$};
\draw (7,1.5) -- (8,1.5); 
\draw[rounded corners] (8,2) rectangle (9,1);
\draw (8.5,1.5) node {$A$};
\draw (9,1.5) -- (9.5,1.5);
\draw (1.5,1) -- (1.5,.5); \draw (3.5,1) -- (3.5,.5); \draw (6.5,1) -- (6.5,.5); \draw (8.5,1) -- (8.5,.5); 
\draw[dotted] (9.5,1.5) -- (10,1.5); 
\end{diagram}\;,
\end{multline}
and simulate the real-time evolution on this finite window using the time-dependent variational principle (TDVP) for site-dependent MPS \cite{jutho2016}. The time evolution can be fully captured by changing $V_i$, as long as the wave packets propagate within the window \cite{milsted2012, zauner2012}.

\noindent\emph{\textbf{The Ising chain}}. %
We illustrate our method by simulating the scattering process between two wave packets in an infinite Ising spin chain defined by the hamiltonian
\begin{equation}
H = \sum_{\braket{ij}} S^z_i S^z_j + \lambda \sum_i S^x_i.\label{eq:isingham}
\end{equation}
In the fully polarized case ($\lambda\to\infty$) the excitations are localized spin flips, and at finite $\lambda$ these become dressed by the quantum fluctuations in the ground state. We work in the symmetric phase ($\lambda>0.5$), but we could similarly study the domain-wall excitations in the symmetry-broken phase with only mild adaptations  (i.e.\ using the two different ground states left and right from the local excitation tensors).
\par
\begin{figure}
	\begin{center}
		\includegraphics[width=\columnwidth]{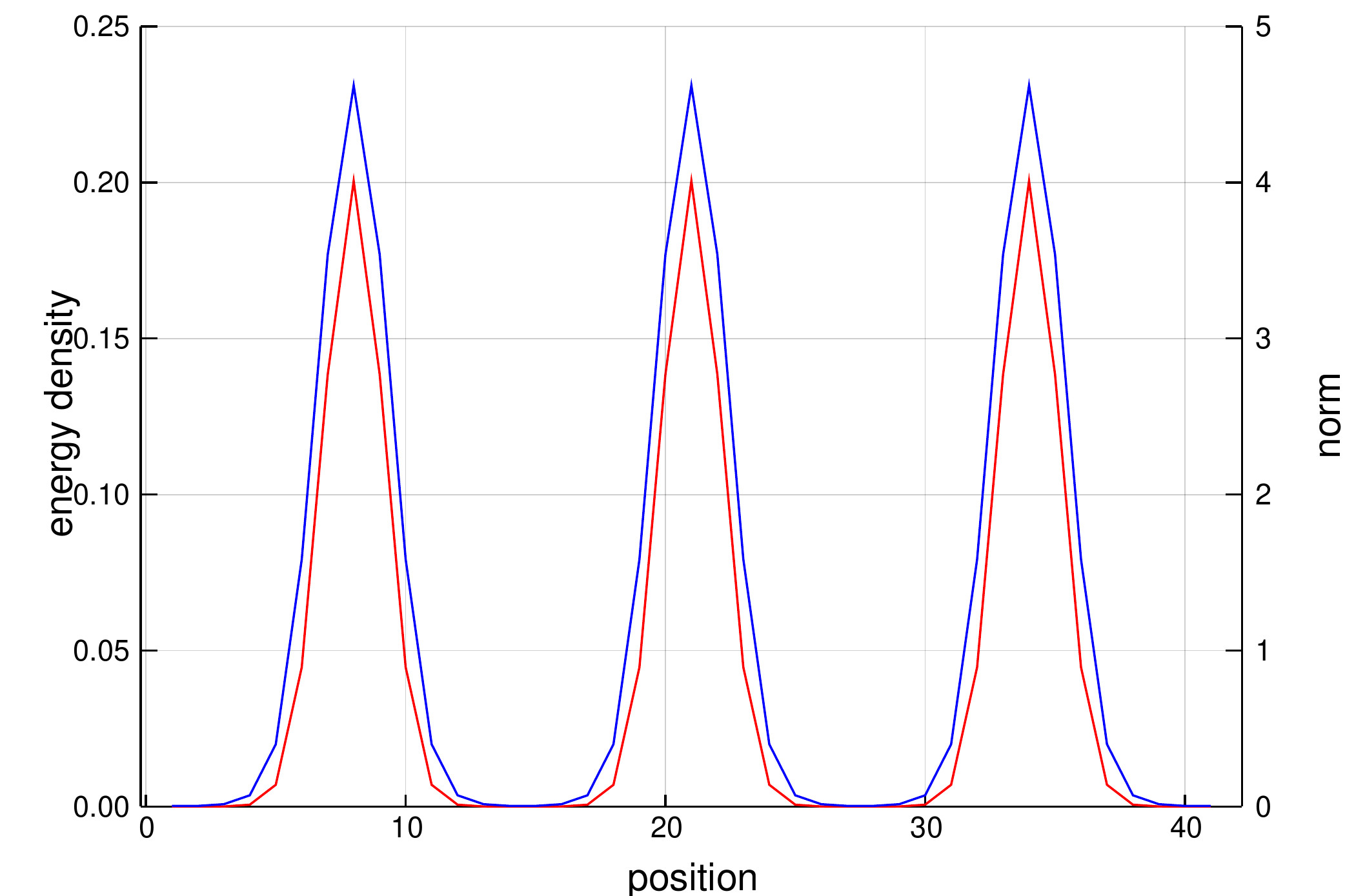}
		\caption{The function $|\sum_{j=1}^{N} f(k_j) B_{k_j} e^{i k_j n}|^2$ (for an Ising chain) evaluated at the position $n$ over several periods $N$ (with here, $N=13$), thus showing the shifted copies (blue). The corresponding excess energy density of the state $\ket{\Phi_f}$ is shown in red.}
		\label{fig:snorm}
	\end{center}
\end{figure}

 We first illustrate the construction of the wave packet. In Fig.~\ref{fig:snorm} we have plotted the norm of the tensor $B_n$ that was optimized as described before, showing the localized wave packet and its shifted copies. In the same figure, we plot the excess energy density (i.e.\ after subtracting the ground state energy density) of the wave function, showing that we have indeed found well-localized packets of energy -- i.e.\ particles -- on top of the correlated ground state. This form of $B_n$ is then truncated with a filter function to keep a single wave packet.

\par
\begin{figure}
	\begin{center}
		\includegraphics[width=\columnwidth]{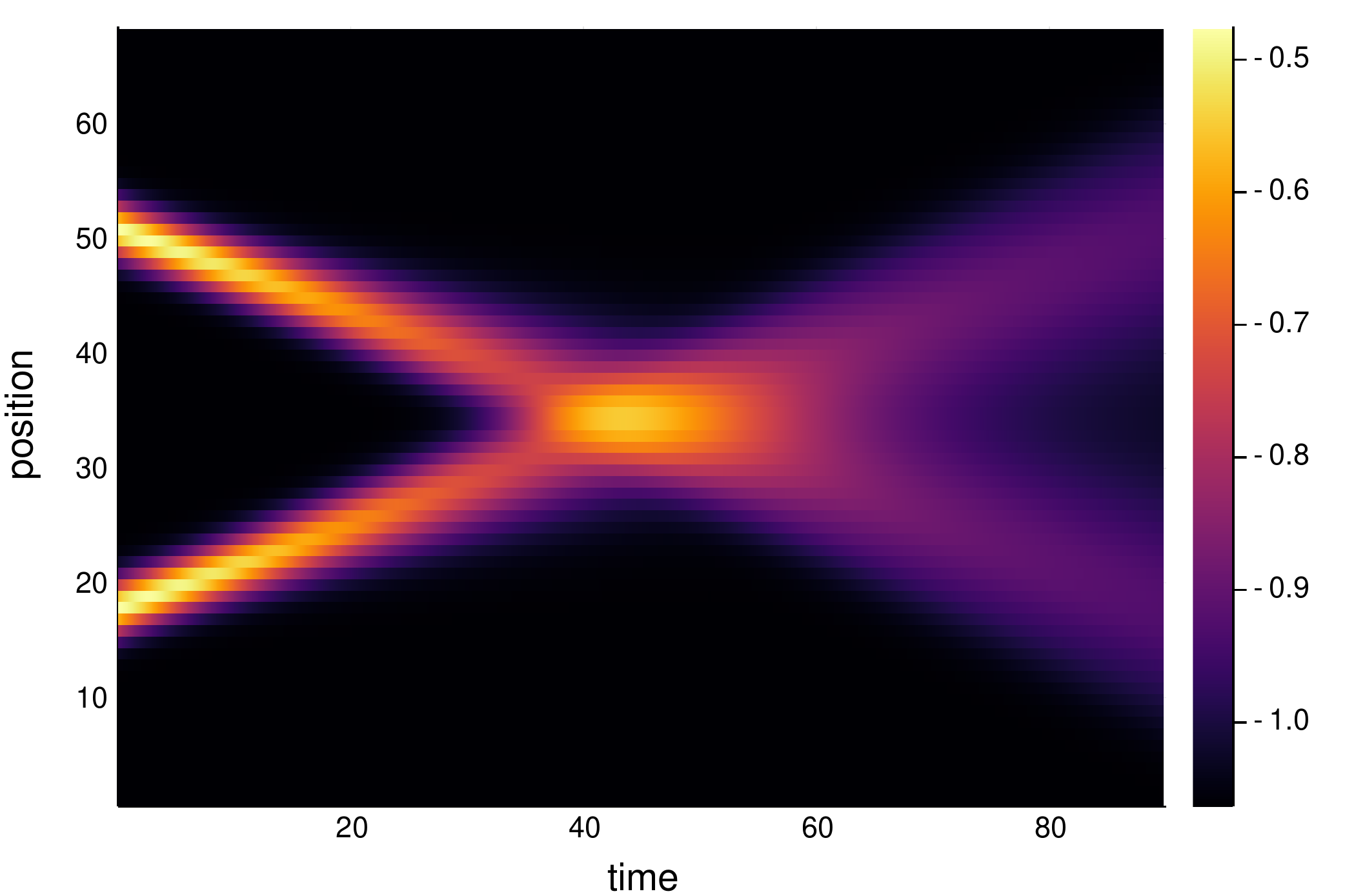}
		\caption{Energy density plot of 2 scattering wave packets in an Ising chain ($\lambda = 1$). We used TDVP \cite{jutho2011} time evolution with time step $dt=0.05$, bond dimension $\chi=20$ and measured the energy density at every point.}
		\label{fig:isingscat}
	\end{center}
\end{figure}

 In Fig.~\ref{fig:isingscat} we display the time evolution of two wave packets that were prepared around momenta $k_0=\pm 1.5$ with a spreading of $v=0.5$. From this figure we can see that the two wave packets nicely propagate through the chain -- there is a small dispersion effect, because of the finite width $v$ of the gaussian -- until they collide and interact. After the collision we find two outgoing wave packets, showing that we have indeed identified stable quasiparticles on top of the correlated ground state. We observe no position shift in the wave packets' trajectory compared to the freely propagating case, which reflects the fact that particles in the Ising chain behave as free fermions \cite{sachdev2011}. To confirm this quantitatively, we have subtracted away the single particle trajectories finding an energy density difference of at most $5 \times 10^{-4}$, in agreement with the expected TDVP error of order $dt^3$.

\begin{figure}
	\begin{center}
		\includegraphics[width=\columnwidth]{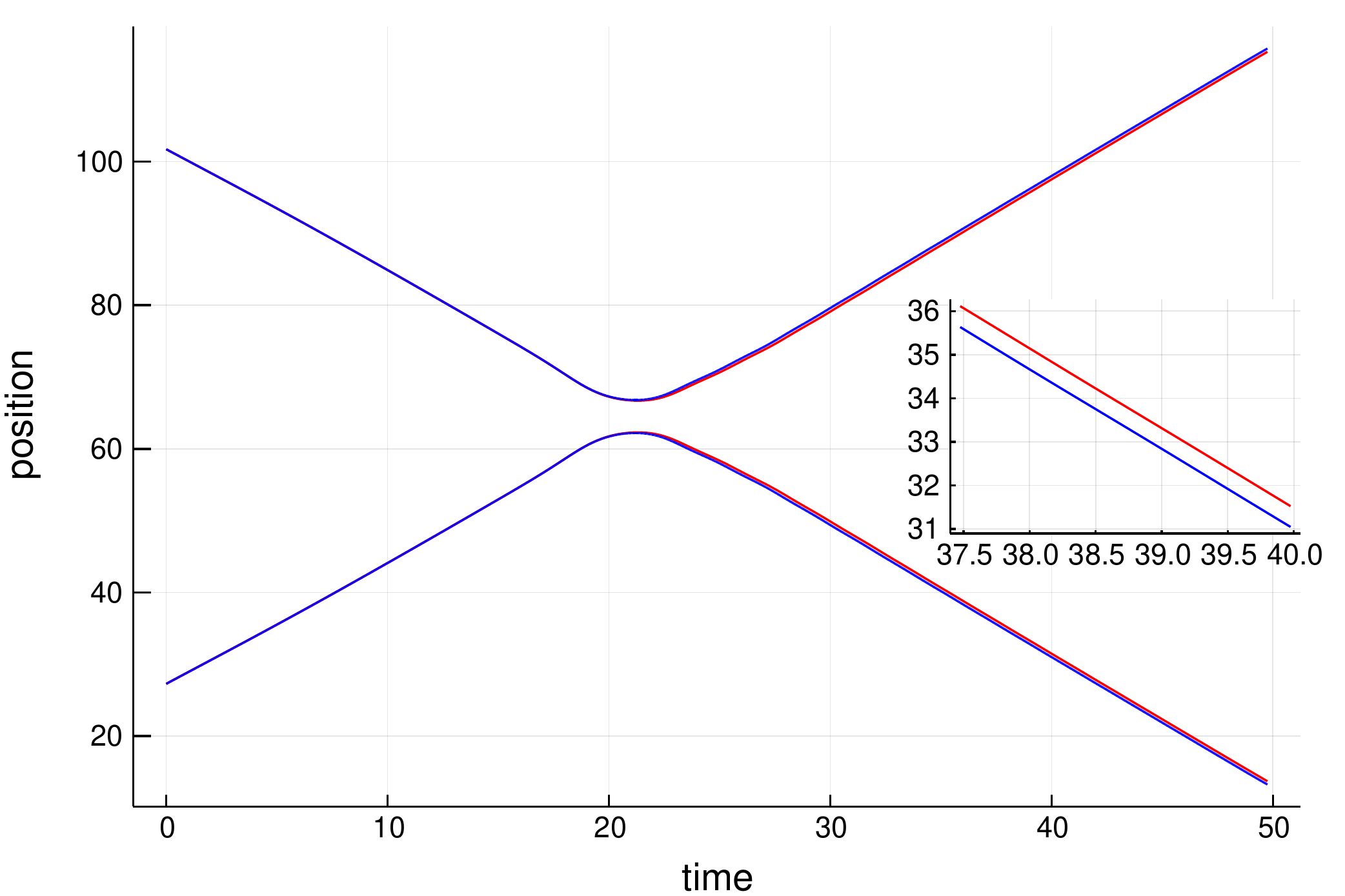}
		\caption{Trajectories of the freely propagating (blue) and colliding wave packets (red) in the spin-1/2 XXZ chain, using TDVP evolution with time step $dt = 0.025$ and bond dimension $\chi = 75$. Inset shows close-up (same axes).}
		\label{fig:weightedhalf}
	\end{center}
\end{figure}

\noindent\emph{\textbf{Spinon-spinon scattering}}. %
Let us now consider the XXZ hamiltonian
\begin{equation}
	H = \sum_i S^x_i S^x_{i+1}+S^y_i S^y_{i+1}+\Delta S^z_i S^z_{i+1},
\end{equation}
which is integrable \cite{Bethe1931,Gaudin2009}. For $\Delta>1$, the elementary excitations are massive particles with fractional spin $s=1/2$ \cite{faddeev1981, Mourigal2013}. The quasiparticle ansatz readily generalizes to this case\cite{jutho2012}, and we find a spinon dispersion relation that agrees with the exact result to high precision. Similarly as before, we create two wave packets with parameters $v = 0.3$ and $k_0=\pm0.7$, and find two well-defined outgoing modes (not shown), a consequence of the fact that particle scattering is purely elastic in integrable systems. Fig.~\ref{fig:weightedhalf} depicts the location of the particles as a function of time, where the approximate position of the wave packet was found by 
\begin{equation} \label{eq:approxpos}
\bar{x} = \frac{1}{N} \sum_{i=0}^{L/2} ie_i^2 , \qquad N = \sum_{i=0}^{L/2} e_i^2,
\end{equation}
where $e_i$ is the excess energy density at site $i$. This can then be compared to the freely propagating case and, in contrast to the Ising case, we find that the collided particles have undergone a displacement. As one can see in the inset of Fig.~\ref{fig:weightedhalf}, we find a displacement that is constant in time with a magnitude $d\approx 0.476$. Using the exact (dressed) scattering phase of excitations on top of the ground state, provided by the integrability framework, the predicted displacement at momenta $k=\pm0.7$ (corrected by calculating the weighted average of the phase shift with two gaussians\cite{suppl}) yields a value of $d=0.4721$, in good agreement with our numerical value. We remark that this constitutes the first direct numerical measurement of the dressed spinon scattering phase shift.

\noindent\emph{\textbf{Magnon-magnon scattering}}. %
Next we study scattering in the non-integrable spin-1 Heisenberg model,
\begin{equation}
H = \sum_i \vec{S}_i \cdot  \vec{S}_{i+1},
\end{equation}
for which it is well known \cite{Haldane1983, White1993} that the elementary excitation is a gapped magnon with spin $s=1$.

The magnon-magnon scattering properties are qualitatively described by the $S$-matrix of the non-linear $\sigma$-model, which was confirmed quantitatively in Refs.~\onlinecite{lou2000,laurens2014}. Because this model is not integrable, two-particle scattering is not purely elastic. An initial state with two particles can also scatter to three- or four-particle state with the same total energy. 

\begin{figure}
	\begin{center}
	\includegraphics[width=\columnwidth]{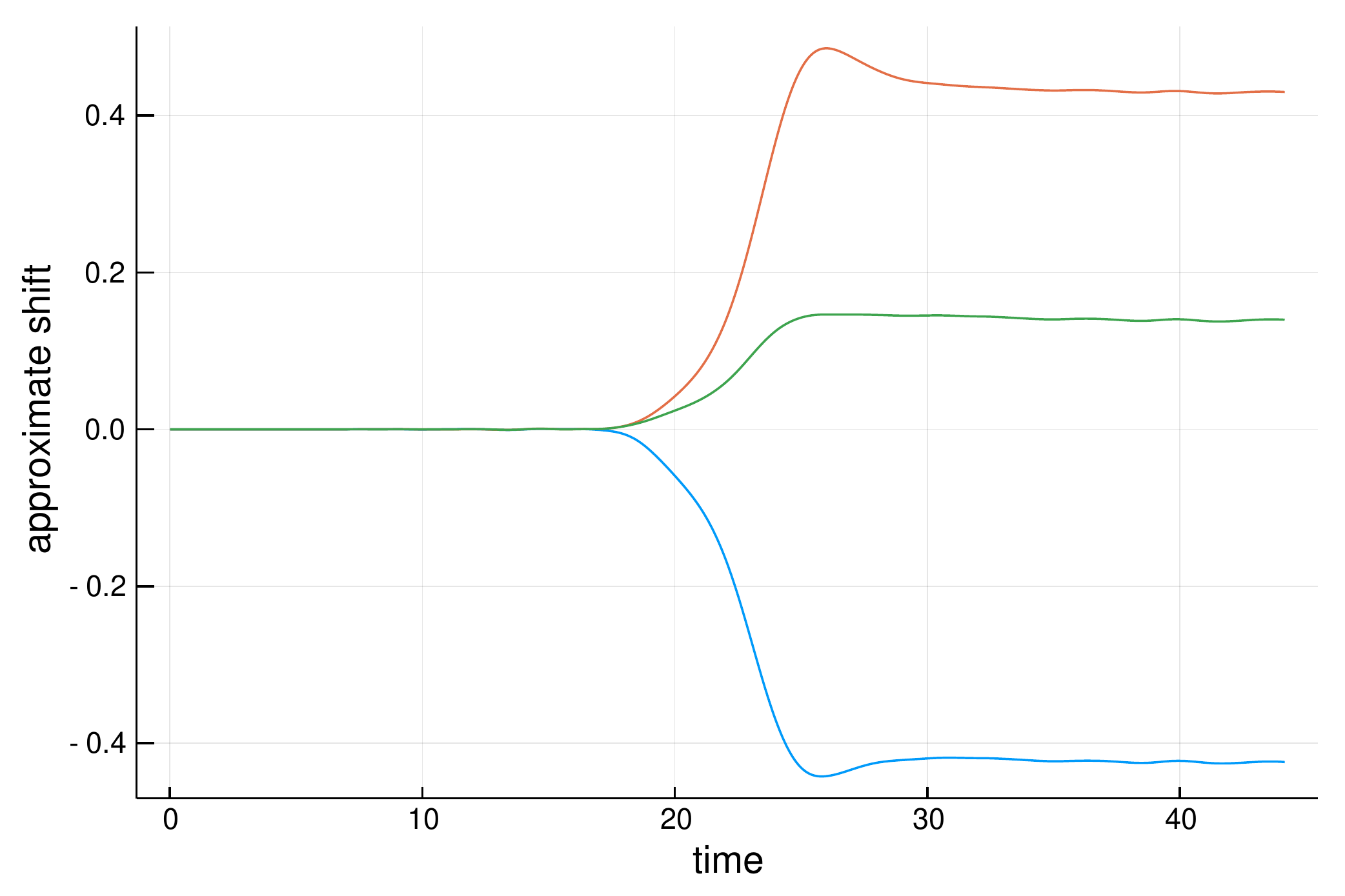}
	\caption{Approximate position shift in the spin-1 Heisenberg chain, for two colliding wave packets with total spin $s=0$ (blue), $s=1$ (orange) and $s=2$ (green).}
	\label{fig:spinone}
	\end{center}
\end{figure}

A two-particle state can have total spin $s=0$, $s=1$ or $s=2$, and we can prepare two-wave-packet states within these three sectors separately. Fig.~\ref{fig:spinone} plots the difference between the approximate position (according Eq.~\eqref{eq:approxpos}) of the colliding and freely propagating wave packets. Even though the model is non-integrable, we observe a well-defined phase shift. We find that the $s=0$ sector is leading as compared the freely propagating particles, whereas the wave packets are lagging in the $s=1$ and $s=2$ sectors. This is in agreement with previously known results for the signs of the scattering lengths in the different sectors \cite{lou2000, laurens2014}.

\noindent\emph{\textbf{Impurity scattering}}. %
We can also describe a wave packet impinging on an impurity in an infinite spin chain using the same techniques. Consider again the spin-1 Heisenberg but with a single spin-2 site at $x=0$ coupling to its neighbors as 
\begin{equation}
H_\mathrm{imp} = \vec{S}_{-1} \cdot \vec{T} + \vec{T} \cdot \vec{S}_{+1} + h T^z,
\end{equation}
where $T^{x,y,z}$ are the spin-2 operators acting at the impurity site and $h$ is a small magnetic field polarizing the impurity spin. We find the ground state of this configuration by optimizing the variational energy within a finite window, embedded in a uniform MPS that corresponds to the ground state of the spin-1 chain.

\begin{figure}
	\begin{center}
		\includegraphics[width=\columnwidth]{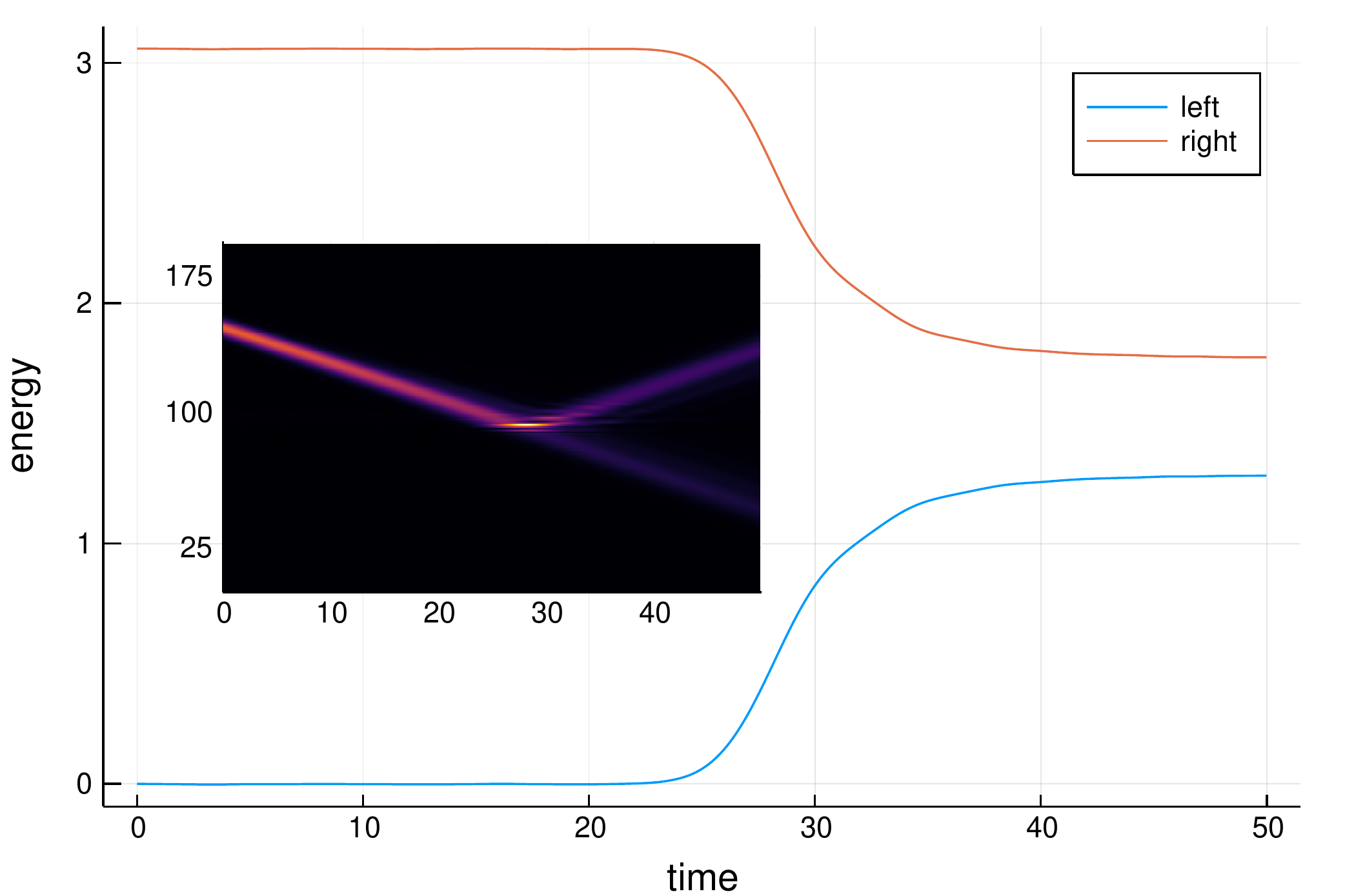}
		\caption{Total energy left and right from the impurity, using TDVP ($dt=0.01$ and $\chi=50$) and $h=0.3$. The inset provides the energy profile for the scattering process.}
		\label{fig:impscatter}
	\end{center}
\end{figure}

\par We have prepared a wave packet, polarized along the $z$, direction with $k_0 = 2.65$ (where the group velocity is negative) and let it impinge on the impurity from the right side. It is interesting to note that, while there may exist local excitations on this impurity spin, these are orthogonal to our initial state if the wave packet is prepared sufficiently far, and will thus remain orthogonal throughout this unitary evolution. Therefore, the wave packet can only reflect on or tunnel through the impurity. In our simulations (see inset of Fig.~\ref{fig:impscatter}), we confirm that no excess in energy stays around the impurity and we find only two propagating wave packets, corresponding to a reflected and transmitted part. By comparing the total energy density left and right from the impurity, we can calculate the transmission and reflection coefficients to be $0.4206$ and $0.5794$ respectively (see Fig.~\ref{fig:impscatter}).

\noindent\emph{\textbf{Outlook}}. %
We have constructed localized wave packets of quasiparticles on top of correlated ground states in generic spin chains, and simulated collision processes in real space and time. We emphasize that our real-space localized wave packets are exceptionally stable through time, against a strongly-correlated background state---this should be compared to the time evolution of a local quench in a system, for which quasiparticles of different energies and momenta lead to an inextricable tangle. We have no knowledge of other approaches that can create these states for generic gapped (non-integrable) systems. Our results for the scattering displacements are in excellent agreement with known analytical results in integrable spin chains. The absence of integrability does not change the fundamental picture as demonstrated by magnon-magnon scattering in the spin-1 Heisenberg model. We have also simulated the scattering of particles off an impurity in the spin chain. 
\par Our results open up the possibility of simulating the low-energy degrees of freedom of generic spin chains and one-dimensional electrons directly in real space and time. In particular, it would be interesting to look at bound-state formation and particle confinement in spin chains, or hole dynamics in electronic chains and ladders. In the context of integrability, we can study the effect of violations of the Yang-Baxter equation and three-particle processes on the real-time dynamics. These results could be used to add extra dissipative terms to the generalized-hydrodynamic equations responsible for decay processes in perturbed integrable systems \cite{Groha_2017} and to compute transport coefficients, related to the dressed scattering shifts\cite{DeNardis2018, Gopalakrishnan2018}, in non-integrable models. Moreover as the dressed scattering shift plays the role of an important input parameter of the non-linear Luttinger liquid field theory \cite{Imambekov2009}, our method allows to extract such data in non-integrable systems.
\par Finally, we believe that our approach can be extended to capture stable localized modes on a representative finite-temperature state; here, we expect to observe a finite lifetime due to multi-particle scatterings with thermally excited particles. Also, starting from the quasiparticle ansatz for two-dimensional systems based on the formalism of projected entangled-pair states \cite{laurens2019b}, we can be apply our approach to scatter quasiparticles in spin and electron systems in two dimensions.

\noindent\emph{Acknowledgements}. %
We would like to thank Boye Buyens, Karel Van Acoleyen and Gertian Roose for inspiring discussions. This work has received funding from the Research Foundation Flanders (G087918N), FWF (BeyondC) and the European Research Council (ERC) under the European Unions Horizon 2020 research and innovation programme (grant agreement No 715861 (ERQUAF) and 647905 (QUTE)). 

\bibliography{./bibliography}

\end{document}


\title{Supplementary material for \\ "Real-time scattering of interacting quasiparticles in quantum spin chains"}

\author{Maarten Van Damme}
\email{maarten.vandamme@ugent.be}
\affiliation{Department of Physics and Astronomy, University of Ghent, Krijgslaan 281, 9000 Gent, Belgium}

\author{Laurens Vanderstraeten}
\affiliation{Department of Physics and Astronomy, University of Ghent, Krijgslaan 281, 9000 Gent, Belgium}

\author{Jacopo De Nardis}
\affiliation{Department of Physics and Astronomy, University of Ghent, Krijgslaan 281, 9000 Gent, Belgium}

\author{Jutho Haegeman}
\affiliation{Department of Physics and Astronomy, University of Ghent, Krijgslaan 281, 9000 Gent, Belgium}

\author{Frank Verstraete}
\affiliation{Department of Physics and Astronomy, University of Ghent, Krijgslaan 281, 9000 Gent, Belgium}

\maketitle

In this Appendix we discuss in more detail how to fix the gauge degrees of freedom [Sec.~\ref{sec:fixing}], how to localize the wave packet (and thus fix de phase degree of freedom) [Sec.~\ref{sec:local}] and how to calculate the phase shift in the XXZ chain by means of the Bethe ansatz [Sec.~\ref{sec:phase}].

\subsection{Fixing the gauge freedom}
\label{sec:fixing}

It is usually convenient to first gauge transform MPS representation of the ground state into the mixed gauge
\begin{equation} 
\ket{\Psi(A_L,A_R,A_C)} =  \dots
\begin{diagram}
\draw (0.5,1.5) -- (1,1.5); 
\draw[rounded corners] (1,2) rectangle (2,1);
\draw (1.5,1.5) node (X) {$A_L$};
\draw (2,1.5) -- (3,1.5); 
\draw[rounded corners] (3,2) rectangle (4,1);
\draw (3.5,1.5) node {$A_L$};
\draw (4,1.5) -- (5,1.5);
\draw[rounded corners] (5,2) rectangle (6,1);
\draw (5.5,1.5) node {$A_C$};
\draw (6,1.5) -- (7,1.5); 
\draw[rounded corners] (7,2) rectangle (8,1);
\draw (7.5,1.5) node {$A_R$};
\draw (8,1.5) -- (9,1.5); 
\draw[rounded corners] (9,2) rectangle (10,1);
\draw (9.5,1.5) node {$A_R$};
\draw (10,1.5) -- (10.5,1.5);
\draw (1.5,1) -- (1.5,.5); \draw (3.5,1) -- (3.5,.5); \draw (5.5,1) -- (5.5,.5);
\draw (7.5,1) -- (7.5,.5); \draw (9.5,1) -- (9.5,.5); 
\end{diagram} \dots,
\end{equation}
This gauge prescription is described in various MPS references, e.g. Ref.~\onlinecite{laurens2019}: $A_L$ and $A_R$ are left and right isometries, which means that the leading respectively left and right eigenvector of their corresponding transfer matrix is the identity matrix. The tensors $A_L$ and $A_R$ are related by a gauge transform $C$, such that $A_L C = C A_R = A_C$. With this gauge transform we can see that the norm of our state reduces to the norm of $A_C$, and thus of $C$, where the singular values of $C$ correspond to the Schmidt coefficients of the state for an arbitrary bipartition. 
\par
\begin{figure}
	\begin{center}
		\includegraphics[width=\columnwidth]{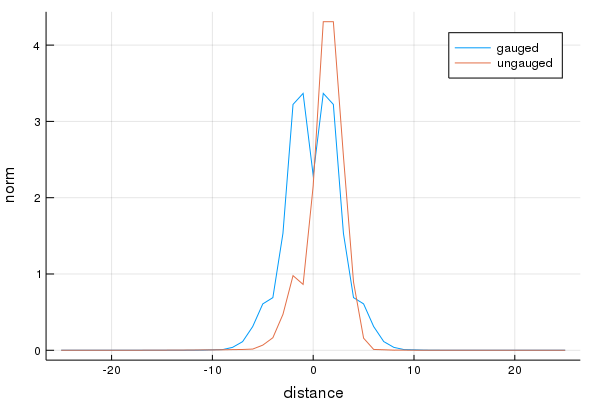}
		\caption{Norm of $B_n$ with and without the new symmetric gauge.}
		\label{fig:asgauge}
	\end{center}
\end{figure}

 The excitation ansatz then becomes :
\begin{equation} 
\ket{\Phi_k(B)} = \sum_{n} \e^{ikn}
\begin{diagram}
\draw (0.5,1.5) -- (1,1.5); 
\draw[rounded corners] (1,2) rectangle (2,1);
\draw (1.5,1.5) node (X) {$A_L$};
\draw (2,1.5) -- (3,1.5); 
\draw[rounded corners] (3,2) rectangle (4,1);
\draw (3.5,1.5) node {$A_L$};
\draw (4,1.5) -- (5,1.5);
\draw[rounded corners] (5,2) rectangle (6,1);
\draw (5.5,1.5) node {$B$};
\draw (6,1.5) -- (7,1.5); 
\draw[rounded corners] (7,2) rectangle (8,1);
\draw (7.5,1.5) node {$A_R$};
\draw (8,1.5) -- (9,1.5); 
\draw[rounded corners] (9,2) rectangle (10,1);
\draw (9.5,1.5) node {$A_R$};
\draw (10,1.5) -- (10.5,1.5);
\draw (1.5,1) -- (1.5,.5); \draw (3.5,1) -- (3.5,.5); \draw (5.5,1) -- (5.5,.5);
\draw (7.5,1) -- (7.5,.5); \draw (9.5,1) -- (9.5,.5);
\draw (1.5,0) node {$\dots$}; \draw (3.5,0) node {$s_{n-1}$}; \draw (5.5,0) node {$s_{n}$}; \draw (7.5,0) node {$s_{n+1}$}; \draw (9.5,0) node {$\dots$};
\end{diagram} \;.
\end{equation}
We can see that $\ket{\Phi_k(B')}$ with $B' = B + A_L X - e^{ik} X A_R$ would represent the same physical state as $\ket{\Phi_k(B)}$. This can be used to greatly simplify the overlap between two momentum states. Consider picking $X$ such that
\begin{equation}
\begin{diagram}
\draw[rounded corners] (1,2) rectangle (2,1);
\draw (1.5,1.5) node (X) {$B'$};
\draw (0,0.5) edge[out=90,in=180] (1,1.5);
\draw (0,-0.5) -- (0,0.5);
\draw (0,-0.5) edge[out=-90,in=180] (1,-1.5);
\draw[rounded corners] (1,-2) rectangle (2,-1);
\draw (1.5,-1.5) node (X) {$A_L$};
\draw (1.5,1) -- (1.5,-1);
\draw (2,1.5) -- (2.5,1.5);
\draw (2,-1.5) -- (2.5,-1.5);
\draw (3.5,0) node (X) {$ = 0$};
\end{diagram},
\end{equation}
such that the overlap between $\braket{\Phi_{k_1}(B_1) | \Phi_{k_2}(B_2)}$ reduces to $\mathrm{Tr}(B_1^\dagger B_2)$. It is this gauge transform that is used when minimizing the energy, because instead of solving a generalized eigenvalue problem, the effective norm becomes the identity and we obtain a regular eigenvalue problem. This gauge is also asymmetric: consider the wavepacket we obtain in spin-1 heisenberg when we follow the procedure from this paper without picking a new gauge, as shown in Fig.~\ref{fig:asgauge}.

A simple way of obtaining a more symmetric gauge is by minimizing the matrix norm of
\begin{equation}
\begin{diagram}
\draw[rounded corners] (1,2) rectangle (2,1);
\draw (1.5,1.5) node (X) {$B'$};
\draw[rounded corners] (1,-2) rectangle (2,-1);
\draw (1.5,-1.5) node (X) {$A_C$};
\draw (1.5,1) -- (1.5,-1);
\draw (2,1.5) -- (2.5,1.5);
\draw (2,-1.5) -- (2.5,-1.5);
\draw (0.5,1.5) -- (1,1.5);
\draw (0.5,-1.5) -- (1,-1.5);
\draw[rounded corners] (-0.5,2) rectangle (0.5,1);
\draw (0,1.5) node (X) {$A_L$};
\draw[rounded corners] (-0.5,-2) rectangle (0.5,-1);
\draw (0,-1.5) node (X) {$A_L$};
\draw (-0,1) -- (-0,-1);
\draw (-1,1.5) -- (-0.5,1.5);
\draw (-1,-1.5) -- (-0.5,-1.5);
\draw[rounded corners] (2.5,2) rectangle (3.5,1);
\draw (3,1.5) node (X) {$A_R$};
\draw[rounded corners] (2.5,-2) rectangle (3.5,-1);
\draw (3,-1.5) node (X) {$A_R$};
\draw (3,1) -- (3,-1);
\draw (3.5,1.5) -- (4.0,1.5);
\draw (3.5,-1.5) -- (4.0,-1.5);
\end{diagram}
\end{equation}
with respect to $X$. This gauge is chosen in such a way that the influence of $B'$ on the virtual space is as local as possible. When the original ground state is reflection invariant, we would therefore also expect the optimal $B'$ to be symmetric.

\subsection{Localizing the wavepacket}
\label{sec:local}

\begin{figure}
	\begin{center}
		\includegraphics[width=\columnwidth]{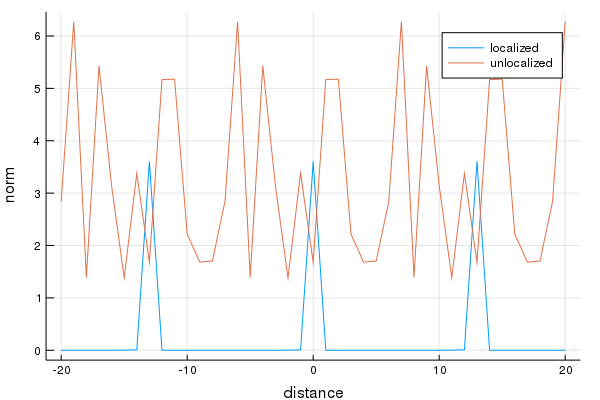}
		\caption{The norm of $|\sum_{j=1}^N f(k_j) B_{k_j} e^{ixk_j}|$ with $f(k_j)=1$ (unlocalized) and with the $f(k_j)$ found in the algorithm (localized), for $N = 2L+1 = 13$.}
		\label{fig:localization}
	\end{center}
\end{figure}

Consider the signal 
\begin{equation}
B_n = \sum_{j=1}^{2L+1} f(k_j) B_{k_j} e^{i k_j n},
\end{equation}
shown in Fig.~\ref{fig:localization}. Because we have uniformly sampled $k_j$ as $k_j = \frac{2\pi}{(2L+1)}j$, $B_n$ is periodic in $n$ with period $2L+1$. In the slice $[-L,L]$ we want the signal to be mostly zero except in a region $[-\Lambda,+\Lambda]$ near $n=0$, so that we can truncate away the copies. Hereto, we write
\begin{equation}
\sum_{|L-n|<L - \Lambda} \left| \sum_{j=1}^{2L+1} f(k_j) B_{k_j} e^{i k_j n} \right|^2 = F^\dagger N_\Lambda F.
\end{equation}
where the vector $F$ has entries $F_j=f(k_j)$. Minimizing this cost function for given $\Lambda$ amounts to finding the vector corresponding to the minimal eigenvalue of the positive definite matrix $N_\Lambda$, which is very efficient. We then pick the solution $F$ for which $\Lambda$ is as small as possible while the cost function is still less then some threshold $\epsilon$, which was set to $\epsilon=10^{-5}$ for the results in this paper. The resulting $B_n$ is then well localized around $n=0$ (and the translated copies thereof), as illustrated in Fig.~\ref{fig:localization}.

\subsection{Spinon scattering shift}
\label{sec:phase}

The lowest lying excitation on top of the ground state of the XXZ spin $1/2$ chain at half-filling is given by the one-spinon excitation \cite{faddeev1981,Mourigal2013}. This excitation can be derived exactly by Bethe ansatz. The logic is to construct the ground state of the chain with a sea of Bethe ansatz excitations (which are magnons) and then perform a single-particle excitation on top of that \cite{Caux2008}. The interacting nature of the system is such that all the others particles in the ground state are shifted by a fraction of the system size, causing a so-called dressing for the particle dispersion relation and scattering shift. Such a dressing effect can be computed exactly by thermodynamic Bethe ansatz \cite{Takahashi1999}. We here report the main results. The (dressed) energy $\varepsilon$ and momentum $k$ of a single spinon is parametrised by the quasi-momentum $\lambda \in [-\pi/2,\pi/2]$ and it is given by 
\begin{align}
\varepsilon(\lambda) & = \pi \sinh \eta \ \mathfrak{s}(\lambda),  \\
k(\lambda) & = 2 \pi \int_{-\pi/2}^\lambda   \mathfrak{s}(\lambda),
\end{align}
with $\mathfrak{s}(\lambda) = \sum_{n=-\infty}^\infty \frac{e^{2 i n \lambda}}{\cosh  n \eta}$ and 
where the parameter $\eta$ is related to the XXZ anisotropy as $\Delta = \cosh \eta$. therefore the group (effective) velocity of the spinon excitation as function of momentum $q$ is given by 
\begin{equation}
v^{\rm eff}(q)= \frac{ \varepsilon'(\lambda)}{k'(\lambda)}\Big|_{\lambda = \lambda(q)}.
\end{equation}

The scattering shift $K^{\rm dr} $ of a spinon excitation is similarly dressed by the non-trivial background of the ground state sea of particles. This is given by the action of the dressing linear operator on the bare scattering shift
\begin{equation}
K^{\rm dr} = [1 + K^{\rm bare} \rho]^{-1} K^{\rm bare}, 
\end{equation}
where the bare scattering shift of Bethe ansatz magnons is $K(\lambda) = \sinh \eta/(\pi( \cosh \eta - \cos (2\lambda)))$ and the distributions of Bethe roots in the ground state is given by $\rho(\lambda) =  \mathfrak{s}(\lambda)/\pi$. The above equation should be intended as the action of a linear operator on a function, where $1$ denotes the identity operator (Dirac delta function). This can be written as a linear integral equation, the solution of which can be obtained via Fourier transform. We then obtain
\begin{equation}
K^{\rm dr}(\lambda ) = \sum_{n=-\infty}^\infty \frac{e^{2 i n  \lambda } }{1 + e^{2 |n| \eta}}.
\end{equation}
Therefore 
 the scattering of two wave-packets with momenta $\bar{q}_1$ and $\bar{q_2}$ produces a displacement in the trajectory $\chi_j$ of the $j-$th packet ($j=1,2$) given by 
\begin{equation}
d_j =   \frac{ K^{\rm dr}(\lambda(\bar{q}_1) - \lambda((\bar{q}_2) ) }{k'(\lambda(\bar{q}_j))}.
\end{equation}

\bibliography{./bibliography}